\documentclass[openany,plhyphen,11pt]{article}
\usepackage{color}
\usepackage{amssymb}
\date{ }
\title{Quantum Market Games
}
\author{E. W. Piotrowski\\ Institute of Theoretical Physics,
University of Bia\l ystok,\\ Lipowa 41, Pl 15424 Bia\l ystok,
Poland\\ e-mail: ep@alpha.uwb.edu.pl\\ and J. S\l adkowski
\\ Institute of Physics, University of Silesia, \\ Uniwersytecka
4, Pl 40007 Katowice, Poland \\ e-mail: sladk@us.edu.pl}
\begin{document}
\maketitle
\def\Z{{\bf Z\!\!Z}}
\def\R{{\bf I\!R}}
\def\N{{\bf I\!N}}
\begin{abstract}
We propose a quantum-like description of markets and economics.
The approach has roots in the recently developed quantum game
theory.
\end{abstract}
 \vspace{5mm}

{\it Introduction} One of the biggest scientific revolutions was
caused by the emergence of quantum physics and the consequent
falsification of the idea of the possibility of distinction
between observer and observed phenomenon. Quantum Zeno effect is
the most famous case in point \cite{1}. This fact is often used
(without any reference to quantum theory) by social scientists to
argue that their theories are beyond mathematical description.
Here we would like to propose a quantum-like approach to economics
whose usefulness, we think, can be verified only by further
investigation. The approach has roots in the recently developed
quantum game theory \cite{2}.  For simplicity and clearness of the
exposition, we shall suppose that there are only two assets on the
market.  The first one, denoted by $\$ $, fulfills the role of
money and its unit is 1$\$ $. The second one, denoted by
$\mathfrak{G}$, is distributed in units 1$\mathfrak{G}$ of price
$c$ in $\$ $. We shall also suppose that both assets can be
exchanged in arbitrary ratio. The traders will be numbered by
$k\in \mathbb{N}$, $\mathbb{N}$ being the natural numbers. The
$k$-th trader implements a strategy (called pure) which is denoted
by  $|\psi\rangle_{k}$ and declares that his whole capital
consisting of $s_{k}$ units of $\mathfrak{G}$ and $d_{k}$ monetary
units will  be engaged in market transactions. A player can be
present on the market in the shape of several traders what allows
him to engage various parts his capital in different ways. The
opposite situation is also possible: several traders can form a
coalition. Such a clique can be described as correlated strategies
of different traders. The actual participation of the $k$-th
trader in the market turnover would be decided by an arbiter
$\mathcal{A}$ who considers the data \{$|\psi \rangle _{k}, s_{k},
d_{k}$\} coming from all traders. The role of the arbiter
$\mathcal{A}$  is performed by an appropriate clearinghouse  who
acts in a fully deterministic way according to rules that are
established by the law or/and tradition. The arbiter $\mathcal{A}$
can be identified with the algorithm used for clearing. Market
transactions, that is capital flows, are settled according to the
strategies put forward by the traders $|\psi \rangle _{k}$ and the
algorithm $\mathcal{A}$. In this way a common price $c$ is set for
all transaction in a given turn. If one considers several markets
connected by traders strategies then simultaneous transaction with
different prices are possible (e.g. arbitrage is possible). Note
that the trader is not obliged to know the market rules but such
knowledge is necessary if he or she wants to
act in a rational way. \\

{\it The quantum model of market} Let the real variable $q$
$$q:= \ln c - E(\ln c) \eqno(1)$$ denotes the logarithm of the
price at which the $k$-th player can buy $\mathfrak{G}$ shifted so
that its expectation value in the state $\mid \psi > _{k}$
vanishes. The expectation value of $x$ is denoted by $E(x)$. The
variable $p$
$$p:= E(\ln c) - \ln c \eqno(2) $$ describes the situation of a
player who is supplying the asset $\mathfrak{G}$ according to his
strategy $|\psi\rangle_k$. Supplying $\mathfrak{G}$ can be
regarded as demanding $\$$ at the price $c^{-1}$ in the
$1\mathfrak{G}$ units and both definitions are equivalent. Note
that we have defined $q$ and $p$ so that they do not depend on
possible choices of the units for $\mathfrak{G}$ and $\$ $. For
simplicity we will use such units so $E(\ln c) =0$. The strategies
$|\psi \rangle_{k}$ belong to  Hilbert spaces $H_{k}$. The state
of the game $|\Psi\rangle_{in}:=\sum_k|\psi \rangle_k$ is a vector
in the direct sum of Hilbert spaces of all players, $\oplus _{k}
H_k$. We will define canonically conjugate hermitian operators of
demand $\mathcal{Q}_k$ and supply $\mathcal{P}_k$ for each Hilbert
space $H_{k}$ analogously to their physical counterparts position
and momentum. This can be justified in the following way. Let
$\exp(-p)$ be a definite price, where $p$ is a proper value of the
operator $\mathcal{P}_k$. Therefore, if one have already declared
the will of selling exactly at the price $\exp(-p)$ (the strategy
given by proper the state $|p\rangle_{k}$) then it is pointless
to put forward any opposite offer for the same transaction. The
capital flows resulting from an ensemble of simultaneous
transactions correspond to the physical process of measurement. A
transaction consists in a transition from the state of traders
strategies $|\Psi\rangle_{in}$ to the describing the capital flow
state $|\Psi\rangle_{out}:=\mathcal{T}_\sigma |\Psi\rangle_ {in}$,
where
$\mathcal{T}_{\sigma}:=\sum_{k_d}|q\rangle_{k_d}\phantom{}_{k_d}
   \negthinspace\langle q|+
 \sum_{k_s}|p\rangle_{k_s}\phantom{}_{k_s}
   \negthinspace\langle p|$  is the projective operator  given by
the division $\sigma $ of the set of traders $\{ k\}$ into two
separate subsets $\{k\}=\{k_d\}\cup\{k_s\}$, that is buying at the
price $e^{q_{k_d}}$ and selling at the price $e^{-p_{k_s}}$ in the
round of transaction in question. The role of the algorithm
$\mathcal{A}$ is to determine the division of the market $\sigma$,
the set of price parameters $\{ q_{k_{d}}, p_{k_{s}}\}$ and the
values of capital flows. The later are settled by the distribution
$$\int_{-\infty}^{\ln c}
\frac{{|\langle q|\psi\rangle_k|}^2}{\phantom{}_k
\negthinspace\langle\psi|\psi\rangle_k}dq \eqno(3)$$ which is
interpreted as the probability that the trader $\mid \psi> _{k}$
is willing to buy  the asset $\mathfrak{G}$ at the transaction
price $c$ or lower \cite{3}. In an analogous way the distribution
$$ \int_{-\infty}^{\ln \frac{1}{c}}
\frac{{|\langle p|\psi\rangle_k|}^2}{\phantom{}_k
\negthinspace\langle\psi|\psi\rangle_k}dp \eqno(4)  $$ gives the
probability of selling $\mathfrak{G}$ by the trader $\mid \psi>
_{k}$ at the price $c$ or greater. These probabilities are in fact
conditional because they describe the situation after the the
division $ \sigma $ is completed. \\

 {\it Maximization of the
capital turnover} Every game is specified by its rules so to
illustrate the action of the algorithm $\mathcal{A}$ let us
consider an example of deterministic clearinghouse algorithm.
Suppose that the clearinghouse maximize the capital turnover at a
uniform price $c$. The assumption of unique for all traders price
$$
\forall_{k_d,k_s} \;(\mathcal{Q}_{k_d}+\mathcal{P}_{k_s})|
\Psi\rangle_{out}=0
$$
corresponds to entanglement in quantum theory. On the market where
transaction are made between two traders, the above condition may
be fulfilled only for every pair of traders separately. Let us for
simplicity limit the class of admissible strategies to those whose
amplitudes $\langle q|\psi\rangle_{k}$ ($\langle
p|\psi\rangle_{k}$) are functions of the variable $q$ ($p$) with
compact supports and square-integrable. After fixing the division
$\sigma $ we can rescale the amplitudes of supply and demand so
integrals of the squares of modules of the resulting functions
 $\langle q|\phi\rangle_{k_d}$ and $\langle p|\phi\rangle_{k_s}$
$$
\langle q|\phi\rangle_{k_d}:=\sqrt{d_{k_d}} \frac{\langle
q|\psi\rangle_{k_d}}{\phantom{}_{k_d}
\negthinspace\langle\psi|\psi\rangle_{k_d}} \;\;,\;\; \langle
p|\phi\rangle_{k_s}:=\sqrt{s_{k_s}} \frac{\langle
p|\psi\rangle_{k_s}}{\phantom{}_{k_s}
\negthinspace\langle\psi|\psi\rangle_{k_s}} \eqno(5)$$ measure the
appropriate capital flows. For a transaction at the price $c$, the
maximal capital flow is the lowest number from the possible total
buying and selling
$$
j(\sigma,c):= \min\Bigl\{ \sum_{k_d}\int^{\ln c}_{-\infty}
|\langle q|\phi\rangle_{k_d}|^2dq\;,
c\sum_{k_s}\int_{-\infty}^{-\ln c} {|\langle
p|\phi\rangle_{k_s}|}^2dp \Bigr\} . \eqno(6) $$
The maximal value
is given by the solution of the equation
$$
\sum_{k_d}\int^{\ln c^\ast}_{-\infty} |\langle
q|\phi\rangle_{k_d}|^2dq = c^\ast\sum_{k_s}\int_{-\infty}^{-\ln
c^\ast} {|\langle p|\phi\rangle_{k_s}|}^2dp . \eqno(7)
$$ There is at least one solution of the Eq. (7) because the
functions are continuous. If there are more solutions we may
choose any of them. The division
$\sigma^\ast=({k^\ast_s},{k^\ast_d})$ corresponding to the maximal
capital flow
$$
j(\sigma^\ast,c^\ast)=\max_\sigma j(\sigma,c^\ast)
$$
gives the changes in the amounts of the asset $\mathfrak{G}$
$$
\label{przyr1-hqmm} \frac{1}{c^\ast}\int^{\ln c^\ast}_{-\infty}
|\langle q|\phi\rangle_{k_d^\ast}|^2dq \;\;\;,\;\;\;
-\int_{-\infty}^{-\ln c^\ast} {|\langle
p|\phi\rangle_{k_s^\ast}|}^2dp \eqno(8)
$$
and the money $\$ $
$$
-\int^{\ln c^\ast}_{-\infty} |\langle
q|\phi\rangle_{k_d^\ast}|^2dq \;\;\;,\;\;\;
c^\ast\int_{-\infty}^{-\ln c^\ast} {|\langle
p|\phi\rangle_{k_s^\ast}|}^2dp \eqno(9)
$$
possessed by the traders after clearing the round of transactions.
The Eq. (7) guarantees that the capital changes will be balanced,
so the described game belongs to specially interesting zero sum
class of game that do not generate capital surplus nor demand
capital inflow (we neglect such cost as brokerage etc).
Maximization of the $\mathfrak{G}$ turnover results in different
price $c^\ast$ and capital flows. This asymmetry can be removed by
separate normalization of the strategies in Eq. (5) that is by
replacing the square roots by
$\sqrt{\frac{d_{k_d}}{\sum_{l_{d}}d_{l_d}}}$ and
$\sqrt{\frac{s_{k_s}}{\sum_{l_{s}}s_{l_s}}}$, \cite{4} and
performing calculations for probabilities instead of capital
flows. In this case the assumption of compactness of the supports
is superfluous. Deutsch arguments \cite{4} can be repeated here to
show that the stochastic interpretation of the presented model is
not necessary. Note that the asymmetry justifies presentation of
the supply and demand curves in terms of probabilities  and not
capital flows \cite{3}. Another, more quantum, algorithm can be
defined if the transactions are made with weights  analogous to
Eq. (6) for all possible divisions $ \sigma $ and prices $c$. Such
transactions should be balanced for not realized bids. This
algorithm may exclude some trades from round of transactions. The
transaction operator would resemble the commonly use in quantum
theory scattering matrix:
$$
\mathcal{T}_{\sigma\alpha}:= I
 +\alpha_d\sum_{k_d}|q\rangle_{k_d}\phantom{}_{k_d}
   \negthinspace\langle q|+
 \alpha_s\sum_{k_s}|p\rangle_{k_s}\phantom{}_{k_s}
   \negthinspace\langle p|,$$ where $\alpha_d,\alpha_s\in\mathbb{R}_+$,
   and $I$ is the identity operator in $\sum_k\oplus H_k$. \\

{\it Market as a measuring apparatus} When a game allows a great
number of players in it is useful to consider it as a two-players
game: the trader $|\psi\rangle_{k}$ against the Rest of the World
(RW). (The player RW has a lot in common with a macroscopic
measuring apparatus.) The concrete algorithm $\mathcal{A}$ may
allow for an effective  strategy of RW (for a sufficiently large
number of players the single player strategy should not influence
on the form of the RW strategy). If one considers the RW strategy
it make sense to declare its simultaneous demand and supply states
because for one player RW is a buyer and for another it is a
seller. To describe such situation it is convenient to use the
Wigner formalism \cite{5}. The pseudo-probability $W(p,q)dpdq$ on
the phase space $\{(p,q)\}$ known as the Wigner function is given
by
\begin{eqnarray*}
W(p,q)&:=& h^{-1}_E\int_{-\infty}^{\infty}e^{i\hslash_E^{-1}p x}
\;\frac{\langle
q+\frac{x}{2}|\psi\rangle\langle\psi|q-\frac{x}{2}\rangle}
{\langle\psi|\psi\rangle}\; dx\\
&=& h^{-2}_E\int_{-\infty}^{\infty}e^{i\hslash_E^{-1}p x}\;
\frac{\langle
p+\frac{x}{2}|\psi\rangle\langle\psi|p-\frac{x}{2}\rangle}
{\langle\psi|\psi\rangle}\; dx,
\end{eqnarray*}
where the positive constant $h_E=2\pi\hslash_E$ is the
dimensionless economical counterpart of the Planck constant.
Recall that this measure is not positive definite except for the
stated bellow cases. In the general case the pseudo-probability
density of RW is a countable linear combination of Wigner
functions, $\rho(p,q)=\sum_n w_n W_n (p,q)$,
 $w_n\geq 0$, $\sum_n w_n =1$.
According to Eq. (3) and (4) (see also Ref. [3]) the diagrams of
the integrals of the RW pseudo-probabilities
$$
F_d(\ln c):=\int_{-\infty}^{\ln c} \rho(p={const.},q)dq \eqno(10)
$$ (RW bids selling at $\exp {(-p)}$)\\and
$$
F_s(\ln c):=\int_{-\infty}^{\ln \frac{1}{c}}
\rho(p,q={const.})dp\eqno(11)
$$ (RW bids buying at $\exp{q}$ ) against the argument $\ln c$ may
be interpreted as the dominant supply and demand curves in Cournot
convention, respectively \cite{3}. Note, that due to the lack of
positive definiteness of $\rho $, $F_d$ and $F_s$ may not be
monotonic functions. Textbooks on economics give examples of such
departures from the low of supply (work supply) and low of demand
(Giffen assets) \cite{6}. We will call an arbitrage algorithm
resulting in non positive definite probability densities {\it a
giffen}. \\

{\it Quantum Zeno effect} If the market  continuously measures the
same strategy of the player, say the demand $\langle q|\psi\rangle
$, and the process is repeated sufficiently often for the whole
market, then the prices given by the algorithm $\mathcal{A}$ do
not result from the supplying strategy $\langle p|\psi\rangle $ of
the player. The necessary condition for determining the profit of
the game is the transition of the player to the state $\langle
p|\psi\rangle $ \cite{3}. If, simultaneously, many of the players
changes their strategies then the quotation process may collapse
due to the lack of opposite moves. In this way the quantum Zeno
effects explains stock exchange crashes. Effects of this crashes
should be predictable because the amplitudes of the strategies
$<p|\psi\rangle$  are Fourier transforms of $<q|\psi\rangle$
Another example of the quantum market Zeno effect is the
stabilization of
prices of an asset provided by a monopolist. \\

{\it Eigenstates of $\mathcal{Q}$ and $\mathcal{P}$} Let us
suppose that the amplitudes for the strategies $\langle
q|\psi\rangle_{k}$ or $\langle p|\psi\rangle _{k}$ that have
infinite integrals of squares of their modules, ($\langle
q|\psi\rangle_k\not\in L^2$) have the natural interpretation as
the will of the $k$-th player of buying (selling) of the amount $
d_k$ ( $s_k$) of the asset $\mathfrak{G}$. So the strategy
$\langle q|\psi\rangle_k= \langle q|a\rangle =\delta(q,a)$ means
that in the case of classifying the player to the set $\{k_d\}$,
refusal of  buying cheaper than at $c=e^{a}$ and the will of
buying at any  price equal or above $e^{a}$. In the case of
"measurement" in the set $\{k_d\}$ the player declares the will of
selling at any price. The above interpretation is consistent with
the Heisenberg uncertainty relation. The strategies $\langle
q|\psi\rangle_2=\langle q|a\rangle$ (or $\langle
p|\psi\rangle_2=\langle p|a\rangle$) do not correspond to the RW
behaviour because the conditions $d_2,s_2>0$, if always fulfilled,
allow for unlimited profits (the readiness to buy or sell
$\mathfrak{G}$ at any price). The demand Eq.~(10) and supply Eq.
(11) functions give probabilities of coming off transactions in
the game when the player use the strategy $\langle
p|{const}\rangle$ or $\langle q|{const}\rangle$ and RW, proposing
the price, use the strategy $\rho$. The authors have analyzed the
efficiency of the strategy $\langle q|\psi\rangle_1=\langle
q|-a\rangle$ in a two-player game when RW use the strategy with
squared module of the amplitude equal to normal distribution
\cite{3}. The maximal intensity of the profit \cite{3} is equal to
0.27603 times the variance of the RW distribution function. Of
course, the strategy $\langle p|\psi\rangle_1=\langle
p|0,27603\rangle$ has the same properties. In such games a=0.27603
is a global fixed point of the profit intensity function. This may
explain the universality of markets on which a single client
facing the bid makes up his/hers mind. Does it mean that such
common phenomena have quantal nature? The Gaussian strategy of RW
\cite{7} can be parametrized by the temperature $T=\beta^{-1}$
(see below). Any decrease in profits is only possible by reducing
the variance of RW (i.e.  cooling). Market competition is the
mechanism responsible for risk flow that allows the market to
attain the thermodynamical  balance. A warmer market influences
destructively on the cooler traders and they diminish the
uncertainty of market prices.\\

{\it Adiabatic strategies} Financial mathematics teach us that the
first moments of the random variables $p$ and $q$ measure the
expected profit from one transaction and the second moments
measure the risk \cite{8}. Therefore we will define the observable
$$
H(\mathcal{P}_k,\mathcal{Q}_k):=\frac{(\mathcal{P}_k-p_{k0})^2}{2m}+
                     \frac{m\omega^2(\mathcal{Q}_k-q_{k0})^2}{2},
                     \eqno(12)
$$
where $p_{k0}:=\frac{
\phantom{}_k\negthinspace\langle\psi|\mathcal{P}_k|\psi\rangle_k }
{\phantom{}_k\negthinspace\langle\psi|\psi\rangle_k}$ $\neq
E(\mathcal{P}_k)$, $q_{k0}:=\frac{
\phantom{}_k\negthinspace\langle\psi|\mathcal{Q}_k|\psi\rangle_k }
{\phantom{}_k\negthinspace\langle\psi|\psi\rangle_k}$,
$\omega:=\frac{2\pi}{\theta}$, and call it {\it the risk
inclination operator}, cf Ref. [3]. $ \theta$ denotes the
characteristic time of transaction introduced in the MM model
\cite{3}. The parameter $m>0$ measures the risk asymmetry between
buying and selling positions. Analogies with quantum harmonic
oscillator allow for the following characterization of quantum
market games. The constant $h_E$ describes the minimal inclination
of the player to risk. It is equal to the product of the lowest
eigenvalue of $H(\mathcal{P}_k,\mathcal{Q}_k) $ and $2\theta$.
Note that $2\theta $ is the minimal interval during which it makes
sense to measure the profit \cite{3}. Except the ground state all
the adiabatic strategies
$H(\mathcal{P}_k,\mathcal{Q}_k)|\psi\rangle={const}|\psi\rangle$
are giffens \cite{5}. Future investigations may reveal the
situations foe which the existence of giffens lead to optimal
strategies.

It should be noted here that in a general case the operators
$\mathcal{Q}_k $ do not commute because traders observe moves of
other players and often act accordingly. One big bid can influence
the market at least in a limited time spread. Therefore it is
natural to consider noncommutative quantum mechanics \cite{9}
where one considers
$$ [ x^{i},x^{k}] = i \Theta ^{ik}:=\Theta \epsilon ^{ik}.$$
The analysis of harmonic oscillator in more then one dimensions
\cite{10} imply that the parameter $\Theta $ modifies the constant
$h_E$ $\rightarrow \sqrt{h_E^{2} + \Theta ^{2}} $ and,
accordingly, the eigenvalues of $H(\mathcal{P}_k,\mathcal{Q}_k)$.
This has the natural interpretation that moves performed by other
players can diminish or increase one's inclination to taking risk.
\\

{\it Correlated coherent strategies} We will define correlated
coherent strategies as the eigenvectors of the annihilation
operator $\mathcal{C}_k$ \cite{11}
$$
\mathcal{C}_k(r,\eta):=\frac{1}{2\eta}\Bigl(1+\frac{ir}{\sqrt{1-r^2}}
\Bigr)\mathcal{Q}_k + i\eta\mathcal{P}_k ,
$$
where $r$ is the correlation coefficient  $r\in[-1,1]$, $\eta>0$.
In these strategies buying and selling transactions are correlated
and the product of dispersions fulfills  the Heisenberg-like
uncertainty relation
$\Delta_p\Delta_q\sqrt{1-r^2}\geq\frac{\hslash_E}{2}$ and is
minimal. The annihilation operators $\mathcal{C}_k$ and their
eigenvectors may be parameterized by
$\Delta_p=\frac{\hslash_E}{2\eta}$,
$\Delta_q=\frac{\eta}{\sqrt{1-r^2}}$ i $r$ This leads to following
form of the correlated Wigner coherent strategy
$$
W(p,q)dpdq=\frac{1}{2\pi\Delta_p\Delta_q\sqrt{1-r^2}}\;e^{-\frac{1}{2(1-r^2)}
\bigl(\frac{(p-p_0)^2}{\Delta^2_p}+\frac{2r(p-p_0)(q-q_0)}{\Delta_p\Delta_q}+
\frac{(q-q_0)^2}{\Delta^2_q}\bigr)}dpdq.
$$ They are not giffens. It can be shown, following Hudson
\cite{12}, that they form the set of all pure strategies with
positive definite Wigner functions. Therefore pure strategies that
are not giffens are represented in phase space $\{(p,q)\}$ by
gaussian distributions. \\

{\it Mixed states and thermal strategies} According to classics of
game theory \cite{13} the biggest choice  of strategies is
provided by the mixed states $\rho(p,q)$. Among them the most
interesting are the thermal ones. They are characterized by
constant inclination to risk,
$E(H(\mathcal{P},\mathcal{Q}))={const}$ and maximal entropy. The
Wigner measure for the $n$-th exited state of harmonic oscillator
have the form \cite{5}
$$
W_n(p,q)dpdq=\frac{(-1)^n}{\pi\hslash_E}\thinspace
e^{-\frac{2H(p,q)}{\hslash_E\omega}}
L_n\bigl(\frac{4H(p,q)}{\hslash_E\omega}\bigr)dpdq,
$$ where $L_{n}$ is the $n$-th Laguerre polynomial. The mixed state  $\rho_\beta$
determined by the Wigner measures $W_ndpdq$ weighted by the Gibbs
distribution $w_n(\beta):=\frac{e^{-\beta
n\hslash_E\omega}}{\sum_{k=0}^\infty e^{-\beta k\hslash_E\omega}}$
have the form
\begin{eqnarray*}
\rho_\beta (p,q)dpdq:&=&\sum_{n=0}^\infty w_n(\beta) W_n(p,q) dpdq\\
&=&\frac{\omega}{2\pi}\;x\; e^{-xH(p,q)} \Bigr|_{x=
\frac{2}{\hslash_E\omega}\tanh(\beta\frac{\hslash_E\omega}{2})}
dpdq .\end{eqnarray*}  So it is a two dimensional normal
distribution. It easy to notice that by recalling that
$\frac{1}{1-t}\thinspace e^\frac{xt}{t-1}=\sum_{n=0}^\infty
L_n(x)t^n$ is the generating function for the Laguerre
polynomials. It seems to us that the above distributions should
determine the shape of the supply and demand curves for
equilibrium markets. There is no giffens on such markets. It would
be interesting to investigate the temperatures of equilibrium
markets. In contrast to the traders temperatures \cite{7} which
are Legendre coefficients and measure "trader's qualities" market
temperatures are related to risk and are positive. The Feynman
path integrals may be applied to the Hamiltonian Eq. (12) to
obtain equilibrium quantum Bachelier model of diffusion of the
logarithm of prices of shares that can be completed by the
Black-Scholes formula for pricing European options \cite{14}.\\

{\it Market cleared by quantum computer} When the  algorithm
$\mathcal{A}$ calculating in a separable Hilbert space $H_k$ does
not know the players strategies it must choose the basis in an
arbitrary way. This may result in arbitrary long representations
of the amplitudes of strategies. Therefore  the algorithm
$\mathcal{A}$ should be looked for in the NP (non-polynomial)
class and quantum markets may be formed provided the quatum
computation technology is possible. Is the quantum arbitrage
possible only if there is a unique correspondence between
$\hslash$ and $h_E$? Penrose ideas concerning the thinking process
suggest searching for new physical phenomena (e.g. giffens)  on
markets, where now the clearing algorithms $\mathcal{A}$ are
constructed in a non-computational way \cite{15}. Was the
hypothetical evidence given by Robert Giffen in the British
Parliament \cite{16} the first ever description of quantum
phenomenon? The market exchange mechanism has inspired the so
called transactional interpretation of quantum mechanics
\cite{17}. The commonly accepted universality of quantum theory
should encourage physicist in looking for traces quantum world in
social phenomena.

\end{document}